\documentclass[a4paper,12pt,oneside]{article}
\usepackage[warn]{mathtext}
\usepackage{graphics}
\usepackage{amsmath}
\usepackage[T2A]{fontenc}
\usepackage{latexsym}
\usepackage[utf8x]{inputenc} 
\usepackage[english]{babel}
\usepackage{graphics}

\title{Tolerances in Flux Compression Generator Design: Theory}
\author{S.~Anishchenko, P.~Bogdanovich, A.~Gurinovich\\\small Research Institute for Nuclear Problems of BSU, Bobruiskaya str., 11, 220030, Minsk, Belarus\\\small e-mail: sanishchenko@mail.ru, gurinovich@inp.bsu.by}
\date{}

\begin{document}
\maketitle
\begin{abstract}
A flux compression generator (FCG) is a device generating a high-power electro-magnetic pulse by compressing magnetic flux with the help of explosion. 
The physical properties of components and materials used for FCG manufacturing and available on the market (wires, explosives, liners (armatures), etc.) might differ from those originally used for FCG design.
As a result, the output parameters of the manufactured generator (e.g. maximum output power) could deviate from the pre-calculated values.
Simple formulas are derived to enable evaluation of change in output power delivered to a purely inductive load from a helical flux compression generator in case when the wire insulation thickness or the detonation velocity of explosive or load inductance $L_l$ are changed.
The obtained formulas establish acceptable tolerances for FCG components produced by a third-party manufacturer.
Design requirement is proposed to ensure minimal dependence of the output power delivered to the purely inductive load when its inductance is varied a bit. The proposed approach implies approximate equality of load inductance and the residual FCG inductance at the instant, when the output power is maximal. 
\end{abstract}

\section{Introduction}

The main features of FCGs invented in the 1950s are now well known~\cite{Sakharov1965,Sakharov1966,Pavlovsky1984,Demidov2012}. 
Designing an FCG  for a particular application one can apply the well-developed theoretical models and numerical methods~\cite{Novac1997,Pikar2001,Lileikis1997,Anishchenko2018}. 
However, the physical properties of components and materials used for FCG manufacturing and available on the market (wires, explosives, liners (armatures), etc.) might differ from those originally used for FCG design.

In particular, some parameters influencing FCG operation could neither specified nor even monitored by manufacturers. For example: a coil (inductor) of the helical FCG is manufactured from high-voltage wires, which the manufacturer usually specifies by the cross-section of copper core (diameter of wire conductor $d$) and HV strength, rather than insulation thickness $i_w$. However, to manufacture the identical inductors with equal initial inductances $L_g$, one should  replicate the geometry with some accuracy and variations in wire shape and insulation thickness might be a challenge.
Since the FCG gain is determined by ratio $L_g/L_l$ any variation in FCG geometry could change the output power delivered to the load. 

The similar situation occurs when some parameters of explosive varies. For FCG operation two velocities are important: detonation velocity $v$ and radial velocity of liner expansion.  Depending on type and quality of explosive material, and its density, the above two determine speed of inductance change $dL_g/dt$, thus influencing on the output power.

Since FCG is a fundamentally single-shot system, it is good to have rapid and refined instruments to evaluate system response on variation of input parameters.
This paper we propose and validate the approach for evaluating the vulnerability of FCG output to variations
in input parameters. 
In section~\ref{sec:theory} the explicit relations enabling to evaluate change of output power for the helical FCG, when input parameters (wire insulation thickness, detonation velocity of explosive, load inductance) are varied. 
Section~\ref{sec:inductance} discusses the value of residual inductance. 
The developed approach is applied for evaluation of typical variations of the output power in section~\ref{sec:analysis}.

\section{Output power}
\label{sec:theory}
 
For a helical FCG the dependence on time of its inductance $L_g(t)$ and resistance $R_g(t)$ are usually expressed as quasi-exponential ~\cite{Pavlovsky1984,Demidov2012}.
Therefore, $L_g(t)$ and $R_g(t)$ can be approximately expressed as follows: 
\begin{equation}
\begin{split}
& L_g(t)=L_0\exp(-\alpha t)+L_\delta, \\
& R_g(t)=R_0\exp(-\alpha t), \\
\end{split}
\end{equation}
where $L_0$ are $R_0$ the initial values of FCG inductance and resistance. For a certain length and diameter of a winding section of inductor  both above values are proportional to cross-section of wire i.e. $d^2$, where $d=d_w + 2 \cdot i_w$ is the diameter of wire,   $d_w$ is the diameter of copper core, $ i_w$ is the insulation thickness. Factor  $\alpha$ is proportional to detonation velocity~$v$.
The residual inductance $L_\delta\approx \text{const}$ is discussed in details hereinafter (see also \cite{Pavlovsky1984b,Zharinov1984}).

Time-dependent current in FCG circuit is expressed as following: 
\begin{eqnarray}
\begin{split}
& I(t)=I_0\cdot\frac{L_0+
	L_\delta+L_l}{L_g(t)+L_l}\cdot\exp\Big(\int_0^t\frac{R_g(\tau)}{L_g(\tau)+L_l}d\tau\Big)\\
& =I_0\cdot\frac{L_0+L_\delta+L_l}{L_g(t)+L_l}\cdot\Big(\frac{L_0+(L_\delta+L_l)e^{\alpha t}}{L_0+L_\delta+L_l}\Big)^{R_0/\alpha L_0}\exp\Big(-\frac{R_0t}{L_0}\Big),\\
\end{split}
\end{eqnarray}
where  $I_0$ is the initial current in the FCG circuit, produced by a seeding source.
Using  $I(t)$ one can express the power delivered to the load:
\begin{eqnarray}
\label{P}
\begin{split}
& P(t)=L_lI\cdot\frac{dI}{dt}\\
& =\frac{(L_0+L_\delta+L_l)I_0^2}{2}\cdot\frac{2L_g(t)L_l(L_0+L_\delta+L_l)}{(L_g(t)+L_l)^3}\cdot\Big(1-\frac{R_0}{\alpha L_0}\Big)\Big(\frac{L_g(t)+L_l}{L_0+L_\delta+L_l}\Big)^{2R_0/\alpha L_0}.\
\end{split}
\end{eqnarray}
Power attains its maximal value $P_{max}$ at some instant
\begin{equation}
t_{max}=\ln\Big(\frac{2(L_0-R_0/\alpha)}{L_l+L_\delta}\Big).
\end{equation}
Using $t=t_{max}$ in \eqref{P} and supposing $L_0\gg L_\delta,L_l$ one can obtain 
\begin{equation}
\label{Pmax}
P_{max}=W_0\cdot\frac{8\alpha L_0L_l}{27(L_l+L_\delta)^2}\cdot\Big[\frac{\alpha L_0-R_0}{\alpha L_0-2R_0}\Big]^3\Big[\frac{(3\alpha L_0-2R_0)L_l}{(L_0+L_l)\cdot(\alpha L_0-R_0)}\Big]^{2R_0/\alpha L_0},
\end{equation}
where $W_0\approx\frac{L_0I_0^2}{2}$ is the initial energy in FCG circuit.

Initial energy $W_0$, which is stored in FCG circuit before the explosion, is determined by the current $I_0$ produced by the seeding source and the initial inductance $L_0$. The initial energy delivered to FCG can be the same for different combinations $L_0$ and $I_0$, and this is the first condition we fix for further consideration.

\section{The residual FCG inductance}
\label{sec:inductance}

The residual FCG inductance is an unavoidable value limiting inductance decrease due to explosive flux compression.  
The total inductance of FCG is defined as a ratio of doubled magnetic energy to the current in FCG circuit.  
Magnetic field penetrates inside the metal parts of FCG and FCG inductance attains addition $L_\delta$.
When FCG liner comes in touch with inductor 
this additional inductance cannot disappear immediately.

\begin{figure}[ht]
	\begin{center}
		\resizebox{160mm}{!}{\includegraphics{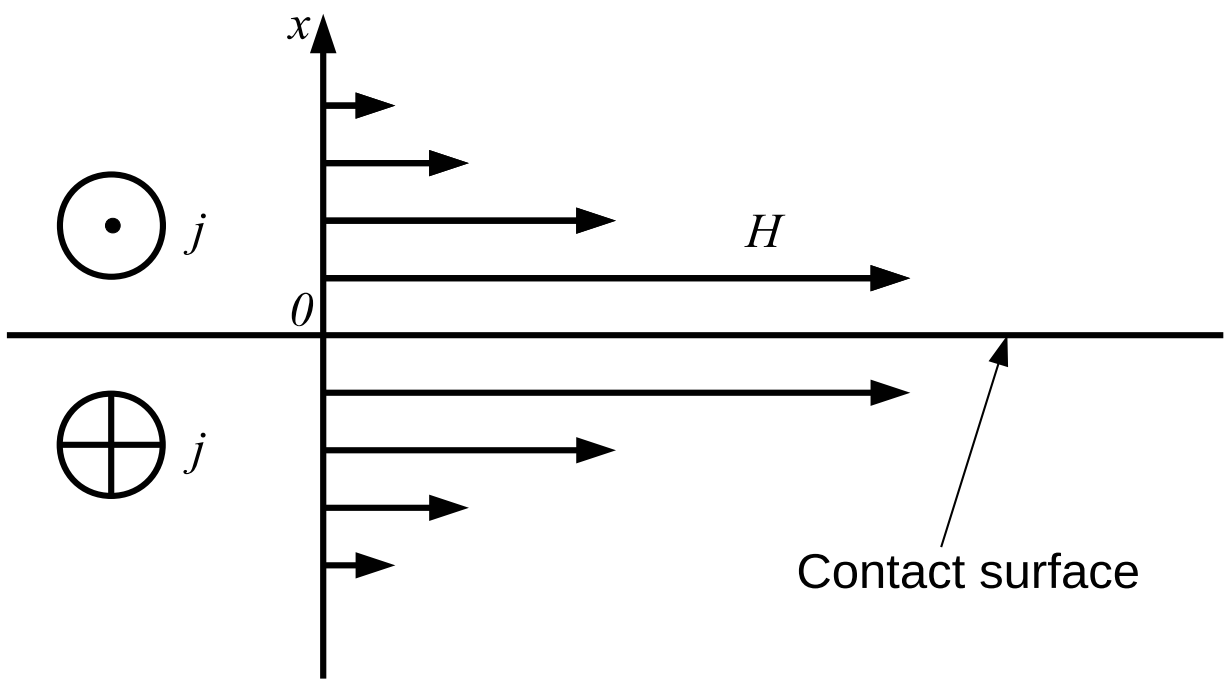}}\\
	\end{center}
	\caption{Two conductors (FCG liner comes in touch with its inductor) and distribution of magnetic field in the vicinity of contact area} \label{fig:conductors}
\end{figure}

Let us consider how magnetic diffusion in FCG induces additional inductance $L_\delta$.
Suppose that magnetic field inside FCG grows exponentially  $H_0e^{t/\tau}$ ($\tau\approx1/\alpha$). The symbol $H_0$ denotes the magnetic field strength on the conductor surface. In FCG, the magnetic field strength $H_0$ can be estimated using winding density $n$ by the formula $H_0\approx nI$.
 
By solving equations of magnetic diffusion one can get distribution of magnetic field inside conductors  (see Fig.~\ref{fig:conductors}) if conductor heating is neglected.

At the instant when conductors carrying the opposite currents approach each other ($t=0$) magnetic field distribution reads as follows: 
\begin{equation}
	\label{InitialField}
	H(x,0)=H_0\exp\bigg(-\frac{|x|}{\delta}\bigg),
\end{equation}
where 
\begin{equation}
	\label{SkinDepth}
	\delta=\sqrt{D\tau}
\end{equation}
is the skin-depth, $D=\sqrt{\eta/\mu_0}$ is the diffusion coefficient, $\eta$ is the specific resistance, $\mu_0=4\pi\cdot10^{-7}$~H/m, $x=0$ is the location of contact area (see Fig.~\ref{fig:conductors}). 

Using Green function
\begin{equation}
	\label{GreenFunction}
	G(x-\xi,t-t_0)=\frac{1}{2\sqrt{\pi D(t-t_0)}}\exp\bigg(-\frac{(x-\xi)^2}{4D(t-t_0)}\bigg)\Theta(t-t_0)
\end{equation}
we can find distribution of field inside metal at $t>0$:
\begin{equation}
	\label{Solution}
	H(x,t)=\int_{-\infty}^{+\infty}G(x-\xi,t)H(\xi,0)d\xi.
\end{equation}
Integral \eqref{Solution} can be expressed as: 
\begin{equation}
	\label{MagneticField}
	H(x,t)=\frac{H_0e^{t/\tau-x/\delta}}{2}\Bigg(1+e^{2x/\delta}-\text{Erf}\bigg(-\frac{-2t/\tau+x/\delta}{2\sqrt{t/\tau}}\bigg)-e^{2x/\delta}\text{Erf}\bigg(\frac{2t/\tau+x/\delta}{2\sqrt{t/\tau}}\bigg)\Bigg),
\end{equation}
where $\text{Erf}$ is the error function. 

\begin{figure}[ht]
	\begin{center}
		\resizebox{80mm}{!}{\includegraphics{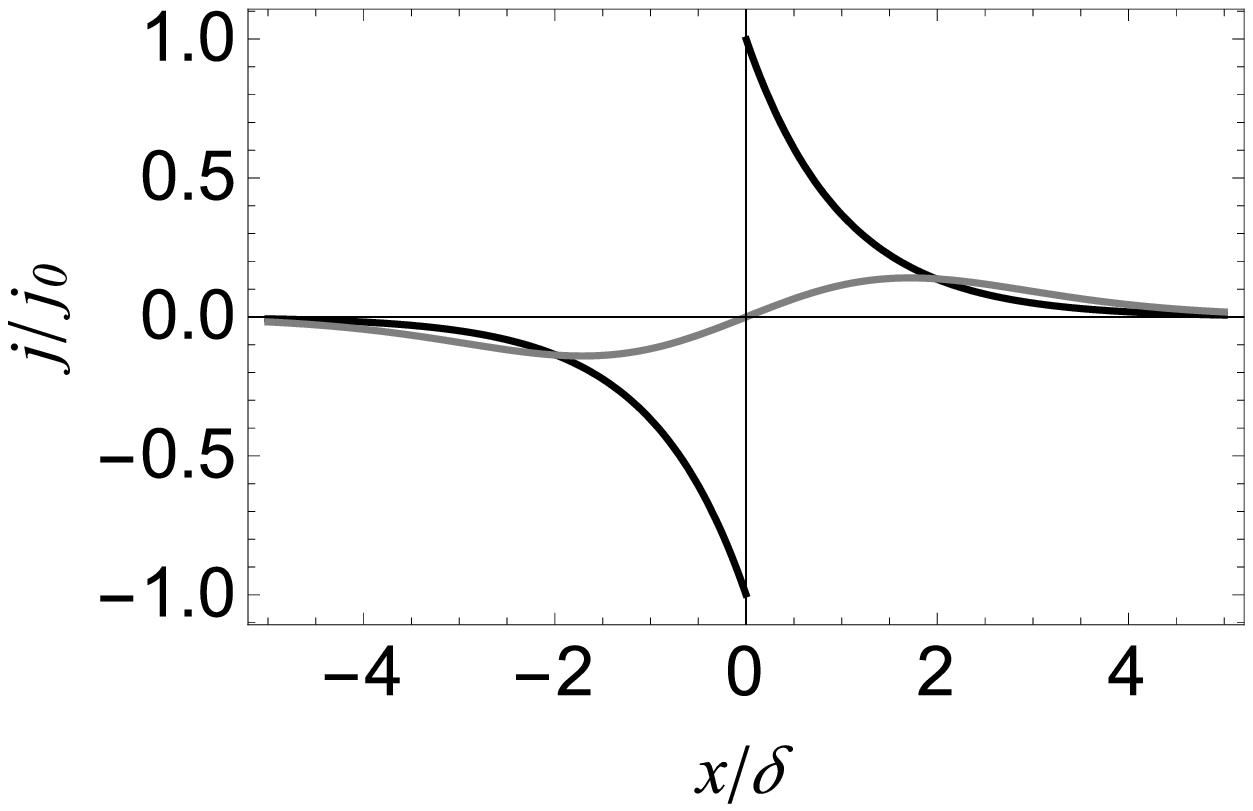}}\resizebox{80mm}{!}{\includegraphics{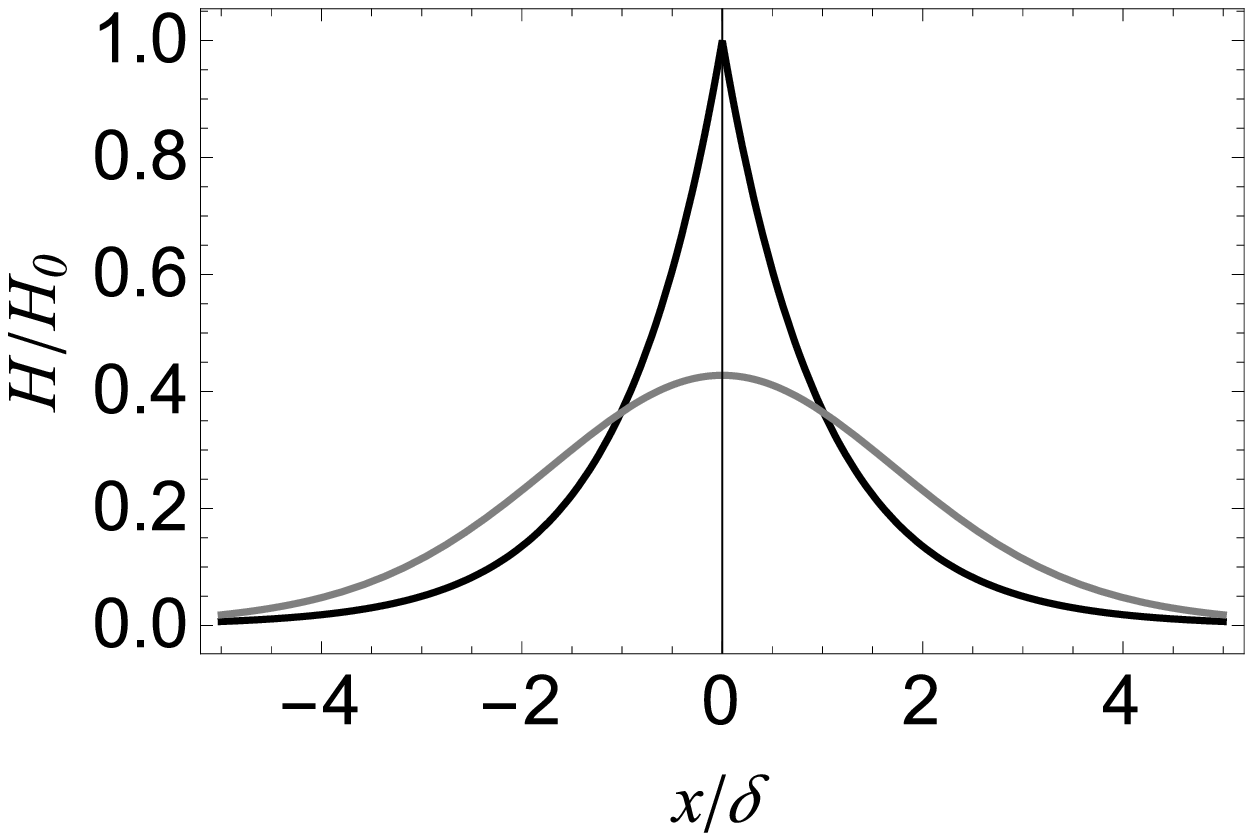}}\\
	\end{center}
	\caption{Distribution of current density and magnetic field in the contact area at $t=0$ (black line) and $t=\tau$ (gray line)} \label{fig:jH}
\end{figure}

Current density $j(x,t)$ inside a conductor is determined by magnetic field $H(x,t)$ as follows:
\begin{equation}
	\label{CurrentDensity0}
	j(x,t)=-\frac{\partial H}{\partial x}.
\end{equation}
Substitution of \eqref{MagneticField} into \eqref{CurrentDensity0} gives
\begin{equation}
	\label{CurrentDensity}
	j(x,t)=\frac {j_0e^{t/\tau-x/\delta}}{2}\Bigg(1-e^{2x/\delta}-\text{Erf}\bigg(-\frac{-2t/\tau+x/\delta}{2\sqrt{t/\tau}}\bigg)+e^{2x/\delta}\text{Erf}\bigg(\frac{2t/\tau+x/\delta}{2\sqrt{t/\tau}}\bigg)\Bigg),
\end{equation}
where $j_0=H_0/\delta$.

Examples of distribution of current density and magnetic field $H(x,t)$ in current carrying conducting parts of FCG are shown in Fig.~\ref{fig:jH}. At the instance $t=0$ current density maximuma are located at $x=0$.  With time they are moving out of the contact area (surface) and at instant $t=\tau$ they appear shifted at distance $2\delta$ out of contact area. Magnetic field at this instant is twice as reduced. 

\begin{figure}[ht]
	\begin{center}
		\resizebox{80mm}{!}{\includegraphics{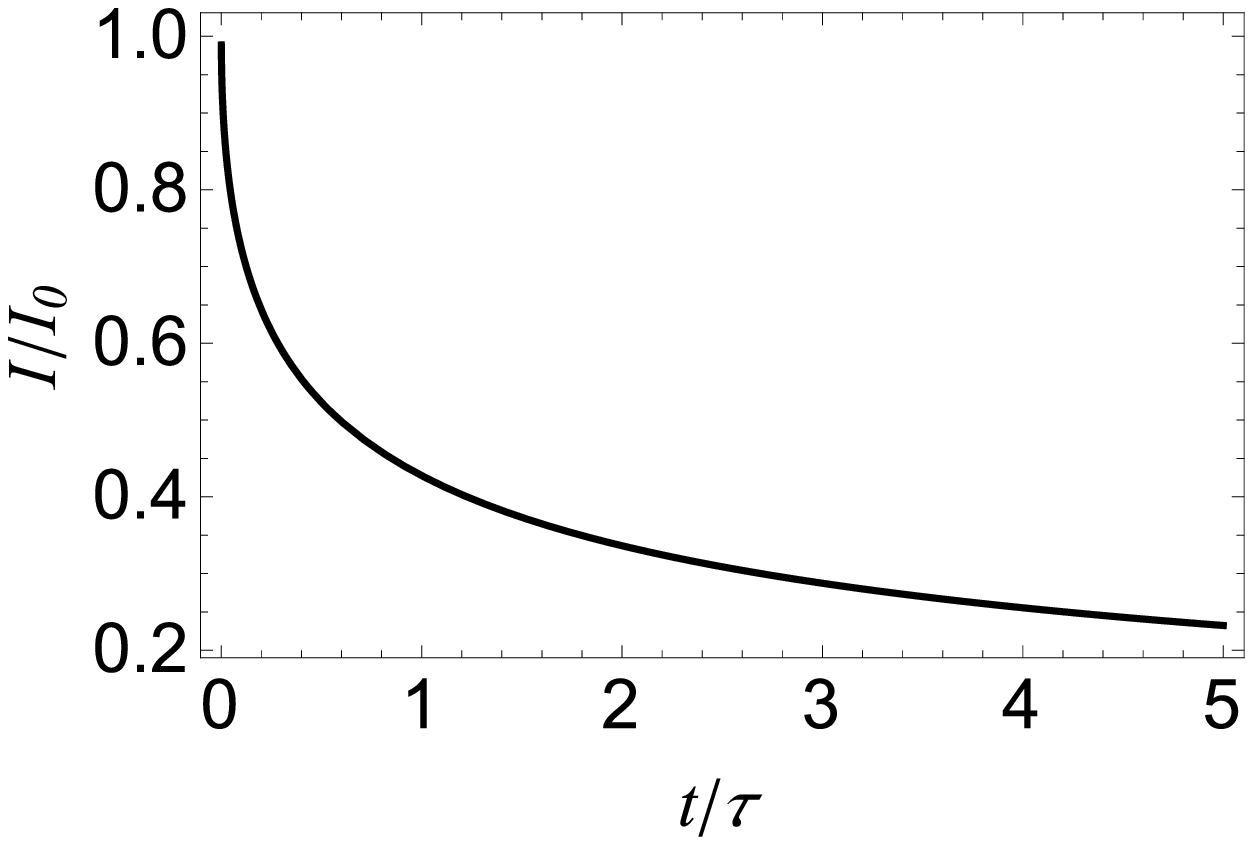}}\resizebox{80mm}{!}{\includegraphics{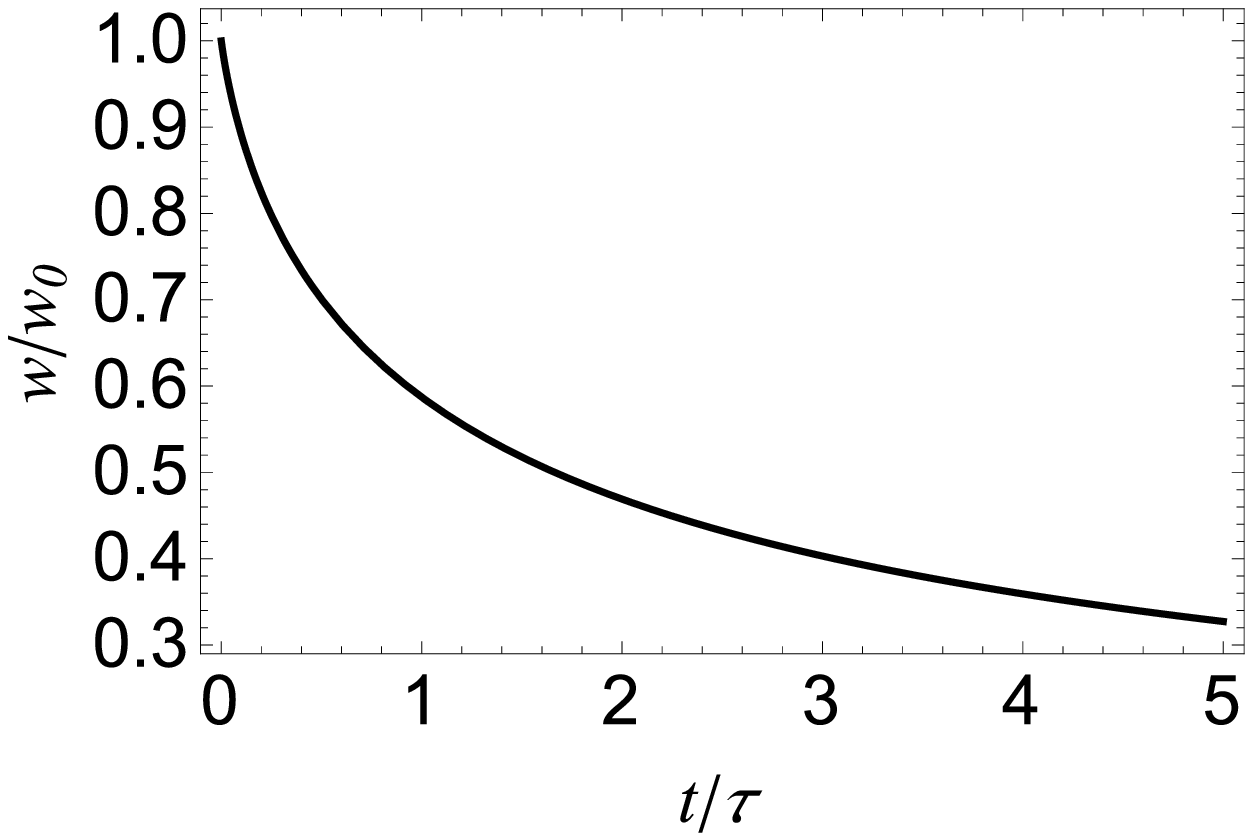}}\\
	\end{center}
	\caption{Reducing of currents (left) and specific magnetic energy (magnetic energy per unit area) (right)} \label{fig:IW}
\end{figure}

Current $I(t)$ in the contact area is proportional to integral $\int_{0}^{+\infty} j(x,t)dx$. Therefore it is convenient to define the typical time of current damping by the following expression:
\begin{equation}
	\label{Current}
	\frac{I(t)}{I_0}=\frac{\int_{0}^{+\infty} j(x,t)dx}{\int_{0}^{+\infty} j(x,0)dx},
\end{equation}
where $I_0=I(0)$.
In similar way we can define the typical damping time for magnetic energy per unit area
 $w(t)=\frac{\mu_0}{2}\int_{0}^{+\infty} H^2(x,t)dx$, 
\begin{equation}
	\label{H}
	\frac{w(t)}{w_0}=\frac{\int_{0}^{+\infty} H^2(x,t)dx}{\int_{0}^{+\infty} H^2(x,0)dx},
\end{equation}
where $w_0=w(0)$.

Normalized current $\frac{I(t)}{I_0}$ and specific magnetic energy $\frac{w(t)}{w_0}$ are shown in Fig.~\ref{fig:IW}. 
The above functions show that current is reduced $e$ times during period

\begin{equation}
	\label{eq:tI}
	t_I\approx1.6\tau,
\end{equation}
while for magnetic energy period of $e$-times damping is longer
\begin{equation}
	\label{eq:tw}
	t_w\approx3.7\tau.
\end{equation}

Typical damping time  $t_{w}$ does not depend on specific conductivity of metal.
It depends only on the speed of magnetic field growth in FCG before the instant when liner and inductor touch each other.  

\begin{figure}[ht]
	\begin{center}
		\resizebox{160mm}{!}{\includegraphics{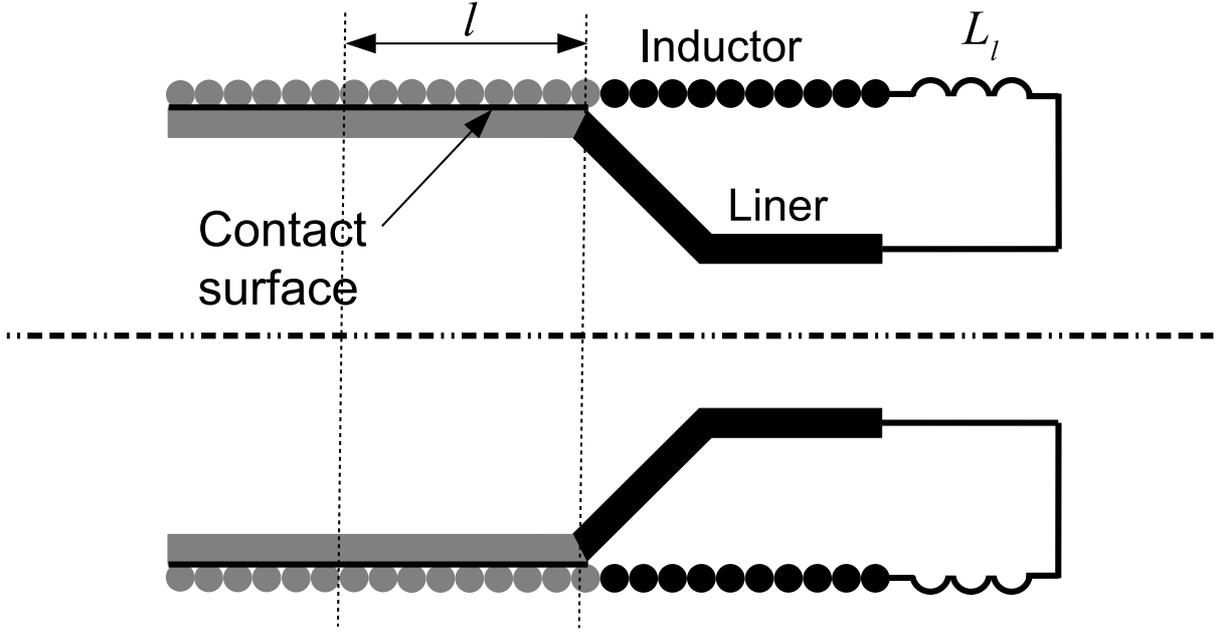}}\\
	\end{center}
	\caption{Contact area (surface) between liner and inductor} \label{fig:hfcg}
\end{figure}

Contact area where magnetic field presents during time $t_{w}$  causes appearance of the residual FCG inductance $L_\delta$, which contributes to the total inductance of FCG.

Let us try to evaluate $L_\delta$ for FCG with moderate parameters. 
The value of magnetic field strength $H_c=34$~MA/m was declared by~\cite{Neuber2005} to be critical  for FCG operation due to nonlinear effects associated with heating of copper. Suppose FCG operates at $H=40$~MA/m.
The energy of magnetic field is equal to product of density of magnetic field $\mu_0H^2/2$ and volume in the contact area
(see Fig.~\ref{fig:hfcg}).
The contact area is a hollow cylinder with the wall thickness equal to depth of magnetic field penetration into conductor (skin-depth $\delta$). Radius of cylinder is equal to inductor radius, length $l$ can be expressed as $l=v \cdot t_{w}$, where $t_{w}$ is the damping time, therefore, volume of this cylinder $V=2\pi Rl\delta$.
According to~\cite{Lewin1964}, the nonlinear effects cause  increase of skin-depth $(1+H/H_c)$ times. 
Therefore, the above expression should be updated as follows: 
\begin{equation}
V=2\pi Rl\delta(1+H/H_c).
\end{equation}
 
Using for calculation $R=0.05$~m, $\tau=7\cdot10^{-6}$~s, $v\approx8\cdot10^3$~m/s, copper resistivity $\eta=1.7\cdot10^{-8}~\Omega\cdot$~m, winding density $n\approx55$~m$^{-1}$,
one gets
$V\approx3.3\cdot10^{-5}$~m$^3$ and $L_\delta\approx\mu_0n^2V\approx125$~nH.

Note that FCG residual inductance $L_\delta$ puts a condition for effective FCG operation: this occurs when  $L_l \approx L_\delta$. If  $L_l\ll L_\delta$ delivery of energy  to load is not effective.

\section{Analysis}
\label{sec:analysis}

Let us analyze the expression (\ref{Pmax}), which gives us dependence of maximal power $P_{max}$ on the above discussed parameters.

For an optimally designed helical FCG $R_0/\alpha L_0\ll1$, therefore two terms braced in square brackets are slowly changing with the change of each other parameter. 
Therefore, the term  $\frac{8\alpha L_0L_l}{27(L_l+L_\delta)^2}\sim\alpha/d^2$ keeps the most of dependence.
Some conclusions follow from it.

Relative change of detonation velocity $\Delta v_{det}/v_{det}$ causes the following relative change in maximal power 
\begin{equation}
\frac{\Delta P_{max}^{(v)}}{P_{max}}\approx\frac{\Delta v_{det}}{v_{det}}.
\end{equation}

Change in wire diameter $\Delta d/d$ results in respective change in maximal power as follows 
\begin{equation}
\frac{\Delta P_{max}^{(d)}}{P_{max}}\approx -2\frac{\Delta d}{d}.
\end{equation}

For varied load inductance $\Delta L_l/L_l$ the response looks like 
\begin{equation}
\frac{P_{max}^{(l)}}{P_{max}}\approx\Big(\frac{L_\delta-L_l}{L_\delta+L_l}\Big)\cdot\frac{\Delta L_l}{L_l}.
\end{equation}

that give us a reason to mention one more conclusion: when $L_l\approx L_\delta$  minor change of load inductance cause no change in maximal power delivered to load.

More exact expressions considering losses in FCG are given by a partial derivative of  $P_{max}$ over a respective parameter.
In this case change of maximal power caused by varied wire diameter reads as follows:
\begin{equation}
\frac{\Delta P_{max}^{(d)}}{P_{max}}\approx -2\Big(1-\frac{2R_0}{\alpha L_0}\Big)\cdot\frac{\Delta d}{d}.
\end{equation}
Variation of detonation velocity contributes to output power as:
\begin{equation}
\frac{\Delta P_{max}^{(v)}}{P_{max}}\approx\Bigg(1+\frac{R_0}{\alpha L_0}\Big[1+2\ln\big(2L_0/3(L_l+L_\delta)\big)\Big]\Bigg)\cdot\frac{\Delta v}{v},
\end{equation}
while change of load inductance $L_l$ results in the following response
\begin{equation}
\frac{\Delta P_{max}^{(l)}}{P_{max}}\approx\Big(\frac{L_\delta-L_l}{L_\delta+L_l}+\frac{2R_0L_l}{\alpha L_0(L_\delta+L_l)}\Big)\cdot\frac{\Delta L_l}{L_l}.
\end{equation}

Let us make evaluations using specific values.
Suppose that $L_\delta\approx L_l$, $R_0/\alpha L_0\ll1$, and wire diameter is $d=2$~mm.
Assume increase of insulation thickness by  $\Delta i_w=0.02$~mm that leads to change in wire diameter $\Delta d=2\Delta i_w=0.04$~mm (i.e 2\%). The associated decrease in output power $\Delta P_{max}^{(d)}/P_{max} \approx 100\%\cdot2\Delta d/d\approx4\%$.

Detonation velocity change from 8 to 7.5~km/s cause decrease of output power $\Delta P_{max}^{(v)}/P_{max} \approx 100\%\cdot\Delta v/v\approx6\%$.

When the load inductance is reduced by 5\%, no visible change in output power could be observed.

\section{Conclusion}

Analytical formulas are derived to enable evaluation of change in output power delivered to a purely inductive load from a helical flux compression generator in case when the wire insulation thickness or the detonation velocity of explosive or load inductance $L_l$ are changed.
The proposed approach establishes acceptable tolerances for FCG components produced by a third-party manufacturer.
Design requirement is proposed to ensure minimal dependence of the output power delivered to the purely inductive load when its inductance is varied a bit: approximate equality of load inductance and the residual FCG inductance is recommended. 

The proposed approach is rather general: any physical quantity can be selected for analysis instead of output power.
Similar formulas can be derived for each of them, for example, maximal energy or current.

\end{document}